\documentclass[12pt]{article}

\usepackage{amsmath}
\usepackage{amsfonts}
\usepackage{times}
\usepackage{epsf}
\def\({\left(}
\def\){\right)}
\def\[{\left[}
\def\]{\right]}
\def\non{ \nonumber }
\def\b{\beta}

\def\a{\alpha}

\begin{document}
\phantom{a}

$$ $$
\centerline{\large\bf New formulae for solutions of quantum }
\centerline{\large\bf
Knizhnik-Zamolodchikov equations on level -4}
\vskip 2cm
\centerline{Hermann Boos
\footnote{on leave of absence from the Institute for High Energy Physics,
Protvino, 142284, Russia}}
\centerline{\it Max-Planck Institut f{\"u}r Mathematik}
\centerline{\it Vivatsgasse 7, 53111 Bonn, Germany}
\vskip 0.5cm
\centerline{Vladimir Korepin}
\centerline{\it C.N.~Yang Institute for Theoretical Physics}
\centerline{\it State University of New York at Stony Brook}
\centerline{\it Stony Brook, NY 11794--3840, USA}
\vskip 0.5cm
\centerline{Feodor  Smirnov
\footnote{Membre du CNRS}}
\centerline{\it LPTHE, Tour 16, 1-er {\'e}tage, 4, pl. Jussieu}
\centerline{\it 75252, Paris Cedex 05, France}
$$ $$
\vskip 1.5cm
\begin{abstract}
We present a new form of solution to the quantum Knizhnik-Zamolodchikov
equation [qKZ] on level $-4$ in a special case corresponding to the
Heisenberg XXX spin chain.
Our form is equivalent to the integral
representation obtained by Jimbo and Miwa in 1996 \cite{JM}.
An advantage of our form is that it is reduced to the product
of single integrals. This fact is deeply related to a cohomological
nature of our formulae. Our approach is also based on the deformation
of hyper-elliptic integrals and their main property -- deformed
Riemann bilinear relation.
Jimbo and Miwa also suggested a nice conjecture which
relates solution of the qKZ on level $-4$ to any correlation
function of the XXX model. This conjecture together with our form
of solution to the qKZ makes it possible to prove a conjecture
that any correlation function of the XXX model can be expressed
in terms of the Riemann $\zeta$-function at odd arguments and
rational coefficients suggested in \cite{bk1,bk2}. This issue
will be discussed in our next publication.
\end{abstract}

\newpage
\noindent
\section{ Introduction.}

This paper originates from the problem of calculation
of correlators in XXX model. Let us remind some history.
First non-trivial correlator on three cites was calculated
by Takahashi \cite{tak}. It happens to be given by $\zeta (3)$
($\zeta$ is Riemann $\zeta$-function). Then tremendous progress
was done by Kyoto group (Jimbo, Miwa, Miki, Nakayashiki) who
provided general formula for correlators in terms of multiple
integrals \cite{jmmn,JM}. Their final results later were confirmed
by Bethe anzatz calculations \cite{Maillet1,Maillet2}. However, the most
interesting result in our opinion consists in relation of certain
generalized correlator introduced by Kyoto group with quantum
Knizhnik-Zamolodchikov equations (qKZ) \cite{FR,SKyo}.

In the special case of level 0 qKZ equations appeared
in the paper \cite{sm1986} in study of form factors for integrable
models of quantum field theory. Kyoto group found that the correlators
are related to the dual case of level -4.

Another source of our inspiration is the papers \cite{bk1,bk2}
in which the conjecture (confirmed by explicit calculations in many
particular cases) was put forward that the correlators in XXX
model can be expressed in terms of values of Riemann $\zeta$-function
with odd positive integer arguments.
This conjecture is rather courageous
because the formula for correlators following from Kyoto group
results are given by multiple integrals which at the first glance
can be expressed as some combination of multiple $\zeta$-function. The claim
that these multiple $\zeta$-values are expressible in terms of single ones
is highly non-trivial. The goal of the present paper is to explain,
at least partly, this miracle.

Let us mention also the paper \cite{bks} in which a generalization
of XXX correlators to inhomogeneous case was considered. It so
happens that in inhomogeneous case the statement concerning
reducibility of correlators becomes much more
transparent. Namely, they are expressed in terms of $\psi$-functions
depending on the inhomogeneity parameters, $\zeta$-values occur in
the homogeneous limit. The logical continuation of the ideas of
this paper requires the consideration of further generalization
of the correlators given by Kyoto group which is related to level -4
qKZ equations.

In the present paper we show the real origin of reducibility of
correlators. Namely, we explain that in the most far reaching
generalization of XXX correlators i.e. in the Kyoto group generalization
the reducibility takes place. To do that we use the relation of level -4
and level 0 qKZ equation. In the latter case the formulae for the
solutions is much nicer \cite{sm1986,book}. The multiple integrals
in these solutions are reduced to single ones from the very beginning.
Using the deformed Riemann bilinear relation \cite{smab} we show that
similar fact is valid for level -4 case.

The fact of reducibility of multiple integrals in solution of
qKZ on level -4 is the Theorem formulated and proved in the Section 5.
Some polynomial coefficients remain undetermined.
The technically  complicated part of the
problem is finding these polynomials. We were able to solve this
problem only partly.

The mathematical meaning of reducibility in question is
illustrated in the Section 6.
The integrals for solutions of qKZ on level -4 can be thought about
as some deformations of integrals of differential forms on affine
Jacobi variety of hyper-elliptic curve. In the classical case the possibility
of reducing the multiple integrals to single ones is explained by the fact
that cohomologies of this variety are especially simple \cite{ns,n}.
>From this point of view one understands why consecutive generalizations
are so useful. The cases of homogeneous XXX, inhomogeneous XXX and Kyoto
generalizations correspond to q-deformation of different Riemann surfaces.
Kyoto generalization corresponds to the case of hyper-elliptic
curve in generic position. Inhomogeneous XXX corresponds to the rational
curve obtained when the  branch points of hyper-elliptic
curve coincide pairwise.
Finally, the homogeneous case corresponds to the situation when all the branch
points come to one point. Obviously, from the point of view of mathematics
one has to consider the less degenerate case.

\section{Jimbo-Miwa solution to qKZ on level -4.}
\vskip 0.5cm
Consider the R-matrix acting in $\mathbb{C}^2\otimes\mathbb{C}^2$ :
\begin{align}
R(\b)=
R_0(\b)\overline R(\b)
\label{R-m}
\end{align}
where
\begin{align}
\overline R(\b)=
\frac {\b +\pi i P}{\b+\pi i}
\label{ovR-m}
\end{align}
$P$ is permutation and
\begin{align}
& R_0(\b)=-\frac
{\Gamma\(\frac \b {2\pi i}\)\Gamma\(\frac 1 2-\frac \b {2\pi i}\)}
{\Gamma\(-\frac \b {2\pi i}\)\Gamma\(\frac 1 2+\frac \b {2\pi i}\)}
\non
\end{align}

The qKZ on level $-4$ are written for a function
$g(\b _1,\cdots,\b _{2n})$ which is meromorphic function of $\b _j$ and
takes values in the tensor product $(\mathbb{C}^2)^{\otimes 2n}$.
We write the qKZ equations \cite{FR,SKyo} close to their
original form which appeared in study of form factors \cite{book}.
Namely, we do not write down indices counting spaces $\mathbb{C}^2$,
for example, we imply that $R (\b _i-\b _j)$ acts in the tensor product
of $i$-th and $j$-th spaces. Also we imply that when the "rapidities"
$\b _i$, $\b _j$ are permuted, corresponding spaces $\mathbb{C}^2$
are permuted as well. With these conventions we can write down the
qKZ equations on level -4 as follows:
\begin{align}
&g(\b _1,\cdots ,\b _{j+1},\b _j,\cdots,\b _{2n})
=R(\b _j -\b _{j+1})
 \ g(\b _1,\cdots ,\b _{j},\b _{j+1},\cdots,\b _{2n})\label{symm}
\\
&g(\b _1,\cdots ,\b _{2n-1},\b _{2n}+2\pi i)
=
g(\b _{2n},\b _1,\cdots ,\b _{2n-1})
\label{Rie}
\end{align}
For application to correlators we need some particular solution
which, according to Jimbo, Miwa \cite {JM} can be written in the form:
\begin{align}
g(\b _1,&\cdots, \b _{2n})=
\non \\
&=\frac 1 {\sum e ^{\b_j}}
\prod\limits _{i<j}\frac 1{\zeta  (\b _i-\b _j)}
\int\limits _{-\infty}^{\infty} d\a _1\cdots \int\limits _{-\infty}^{\infty}
d\a _{n-1}\prod\limits _{i,j}\varphi (\a _i -\b _j)\non\\
&\times
\prod\limits _{i>j} \frac {\sinh
(\a _i-\a _j)}{\a_i-\a _j-\pi i}\ e^{-\sum \a _i +
\frac 1 2 \sum \b_j}
D(\a _1,\cdots \a _{n-1}|\b _1,\cdots ,\b _{2n})\non
\end{align}
where
$$\varphi (\a)=\Gamma \(\frac 1 4 +\frac {\a}{2\pi i}\)
\Gamma \(\frac 1 4 -\frac {\a}{2\pi i}\)$$
$$\zeta (\b)=
\exp\(-\int\limits _{0}^{\infty}
\frac {\sin ^2\frac 1 2(\b+\pi i)k\ e^{-\frac {\pi k}2}}
{k\sinh (\pi k)\cosh \(\frac {\pi k}2\)}\)$$
$D(\a _1,\cdots \a _{n-1}|\b _1,\cdots ,\b _{2n})$ is a polynomial
taking values in $(\mathbb{C}^2)^{\otimes 2n}$. Due to the
symmetry property (\ref{symm}) it is sufficient to give its
$\{-\cdots - +\cdots +\}$-component:
\begin{align}
&D(\a _1,\cdots \a _{n-1}|\b _1,\cdots ,\b _{2n})_{-\cdots - +\cdots +}=
\non\\
&=
\prod\limits _{k}\prod\limits _{j>k}\(\a _k -\b _j+\frac {\pi i} 2\)
\prod\limits _{j<k}\(\a _k -\b _j-\frac {\pi i} 2\)
\non\\
&\times \sum\limits _{l=1}^{n} \(2\sum \a _k+2\b _l-\sum\b_j +\pi i(2l-1)\)
\prod\limits _{j\ge l}
\frac {\a _j -\b _j-\frac {\pi i} 2}{\a _j-\b_{j+1}+\frac {\pi i} 2}
\label{D}
\end{align}
This formula has one not very pleasant feature: it is not symmetric
with respect to $\b _1,\cdots ,\b _n$, the symmetry takes place
only for the integral.
Notice that this solution belongs to the invariant with respect
to action of $SU(2)$ (singlet) subspace of $(\mathbb{C}^2)^{\otimes 2n}$.
However, the main trouble with this formula is in presence of
denominators $\a _r-\a _s-\pi i$ which makes the integrals essentially
multi-fold.

\section{Smirnov solution to qKZ on level 0.}
\vskip 0.5cm
\noindent
Originally qKZ equations appeared for level 0 as form factor equations
\cite{sm1986}.
It is convenient to write them for a covector from $\mathbb{C}^{\otimes 2n}$
denoted by $f(\b _1,\cdots ,\b _{2n})$:
\begin{align}
&f(\b _1,\cdots ,\b _{j+1},\b _j,\cdots,\b _{2n})
=
f(\b _1,\cdots ,\b _{j},\b _{j+1},\cdots,\b _{2n})R(\b _{j+1}-\b _j)
\label{symm0}\\
&f(\b _1,\cdots ,\b _{2n-1},\b _{2n}+2\pi i)
=
f(\b _{2n},\b _1,\cdots ,\b _{2n-1})\label{Rie0}
\non
\end{align}
We need solution belonging to the singlet subspace.
The difference with level -4 case
 seems to be minor, but the formulae for
solutions are much nicer. Many solutions can be written
which are counted by \newline
$\{k_1,\cdots ,k_{n-1}\}$, with $|k_j|\le n-1$, $\forall j$:
\begin{align}
&f^{\{k_1,\cdots ,k_{n-1}\}}(\b _1,\cdots, \b _{2n})=
\prod\limits _{i<j}\zeta (\b _i-\b _j)
\int\limits _{-\infty}^{\infty} d\a _1\cdots \int\limits _{-\infty}^{\infty}
d\a _{n-1}\prod\limits _{i,j}\varphi (\a _i -\b _j)\non\\
&\times
\text{det}|e^{k_i\a _j}|_{1\le i,j\le n-1}\
\ h(\a _1,\cdots \a _{n-1}|\b _1,\cdots ,\b _{2n})\non
\end{align}
where $h$ is skew-symmetric w.r. to $\a$'s polynomial.
The $\{-\cdots -+\cdots +\}$ component of $h$ is given by
\begin{align}
&h(\a _1,\cdots \a _{n-1}|\b _1,\cdots ,\b _{2n})_{-\cdots -+\cdots +}=&\non\\
&
u(\a _1,\cdots \a _{n-1}|\b _1,\cdots ,\b _{n}|\b _{n+1},\cdots ,\b _{2n})
\prod_{j=1}^n\prod_{j'=n+1}^{2n}
\frac{1}{\b_j-\b_{j'}+\pi i}&\non
\end{align}
where
\begin{align}
&u(\a _1,\cdots \a _{n-1}|\b _1,\cdots ,\b _{n}|\b _{n+1},\cdots ,\b _{2n})=
\non\\&=
\text{det}(A_i(\a _j|\b _1,\cdots ,\b _{n}|\b _{n+1},\cdots ,\b _{2n}))
|_{i,j=1,\cdots , n-1}
\end{align}
as for polynomials $A_i(\a)$ which depend on $\b _j$ as on parameters,
it is convenient to write for them generating function:
\begin{align}
&\sum\limits _{i=1}^{n-1}\gamma ^{n-i-1}
A_i(\a|\b _1,\cdots ,\b _{n}|\b _{n+1},\cdots ,\b _{2n})=
\label{gener}\\&=
\frac {\prod\limits _{j=1}^{2n}(\a -\b _j+\frac {\pi i}2)}{\a -\gamma +\pi i}-
\frac {\prod\limits _{j=1}^{2n}(\a -\b _j-\frac {\pi i}2)}
{\a -\gamma -\pi i}+\non\\
&+\frac {\pi i\prod\limits _{j=1}^{n}(\a -\b _j-\frac {\pi i}2)
(\gamma-\b _{n+j}+\frac {\pi i}2)}
{(\a-\gamma)(\a -\gamma -\pi i)}+
\frac {\pi i\prod\limits _{j=1}^{n}(\gamma -\b _j-\frac {\pi i}2)
(\a-\b _{n+j}+\frac {\pi i}2)}
{(\a-\gamma)(\a -\gamma +\pi i)}\non
\end{align}
This expression is manifestly symmetric with respect to two groups
of $\b _j$.
Important difference with the previous case is that
there are
no denominators here, effectively the integral is reduced to one-fold
ones.

There is an explicit formula expressing $h$ in terms of $u$:
\begin{align}
&h(\a _1,\cdots \a _{n-1}|\b _1,\cdots ,\b _{2n})=\non\\&=
\sum\limits
_{\{1,\cdots,2n\}=\{i_1,\cdots ,i_n\}\cup \{j_{1},\cdots ,j_{n}\}}
u(\a _1,\cdots \a _{n-1}|\b _{i_1},\cdots ,\b _{i_n}|
\b _{j_1},\cdots ,\b _{j_n})
\non\\&\times\prod\limits _{p,q=1}^n \frac 1 {\b _{i_p}-\b _{j_q}}
w_{\epsilon _1,\cdots,\epsilon _{2n}}(\b _1,\cdots\b_{2n})\non
\end{align}
where $\epsilon _{i_p}=-$, $\epsilon _{j_p}=+$, and the basis
$w_{\epsilon _1,\cdots,\epsilon _{2n}}(\b _1,\cdots\b_{2n})$ which is
described in \cite{book} satisfies important relation:
\begin{align}
&w_{\epsilon _1,\cdots,\epsilon_i,\epsilon_{i+1},\cdots
\epsilon _{2n}}(\b _1,\cdots\b_i,\b_{i+1},\cdots\b_{2n})
\overline R(\b _{i+1}-\b _i)=
\non\\ &=
w_{\epsilon _1,\cdots,\epsilon_{i+1},\epsilon_i,\cdots
\epsilon _{2n}}(\b _1,\cdots\b_{i+1},\b_i,\cdots\b_{2n})
\end{align}
with $\overline R$ given by the formula (\ref{ovR-m}).
Our main statement is that it is possible to write down similar
formula for level -4 case. But before explaining this point we have to
remind some properties of deformed hyper-elliptic integrals.

\section{Deformed hyper-elliptic integrals.}

In this section we follow mostly the paper \cite{smab}.
The solutions to level 0 qKZ equations are expressed in
terms of the following integrals:
\begin{align}
&\langle P\ |\ p\rangle=
\int\limits _{-\infty}^{\infty}
\prod\limits _j\varphi (\a -\b _j)\ P(e^{\a })\ p(\a )
\ e^{-(n-1)\a }d\a\label{abint}
\end{align}
where $p(\a )$ and $P(e^{\a})$ are polynomials which depend
respectively on $\b _j$ and $e^{\b _j}$ as on parameters. For integral
to converge we have to require $\text{deg}(P)\le 2n-2 $.

We shall intensively use the asymptotic series in $\a ^{-1}$ for the
function $\prod\varphi (\a-\b _j)$. These series (denoted by $\Phi(\a)$)
can be defined from their main property:
\begin{align}
\Phi(\a +2\pi i)=\Phi (\a)
\prod\limits_{j=1}^{2n}\frac {\a -\b _j+\frac {\pi i}2}
{\a -\b _j+\frac {3\pi i}2}
\end{align}
In the integrals (\ref{abint}) we shall never consider analogues of
the differentials of third kind i.e. we shall require:
\begin{align}
&\text{res}_{\a =\infty}\(p(\a)\Phi (\a)\)=0\label{res}
\end{align}
It is easy to see that in this case
\begin{align}
&P(e^{\a})=\biggl(\prod (e^{\a}+ie^{\b _j})-\prod (e^{\a}-ie^{\b _j})
\biggr)e^{-\a}\simeq 0
\label{closed}
\end{align}
i.e. this polynomial gives zero when substituted into the integral.
As for remaining polynomials $p(\a )$ one can show that only $2n-2$
give non-trivial result.
Indeed, with every polynomial $l (\a )$
we can associate  an "exact form":
$$ l(\a +\pi i)\prod \(\a -\b _j +\frac {\pi i} 2\)-
l(\a -\pi i)\prod \(\a -\b _j -\frac {\pi i} 2\)$$
Being put under the integral (\ref{abint}) this "exact form" gives zero.
On the other hand one can reduce degree of any polynomial
adding "exact forms" up to $2n-2$.

For the basis of nontrivial polynomials
we take:
\begin{align}
&s_k(\a  )=A_k(\a |\b _1,\cdots ,\b _n |\b _{n+1},\cdots ,\b _{2n})\non\\
&s _{-k}(\a )=\a ^{n-k-1}, \qquad k=1,\cdots , n-1
\end{align}
Define
$$\Delta (f)(\a)=f(\a+\pi i)-f(\a-\pi i)$$
The following  skew-symmetric
pairing is well defined on polynomials satisfying (\ref{res}):
\begin{align}
&p\circ q=\text{res}_{\a =\infty}\(p(a )\Phi (\a )\Delta ^{-1}(q(\a )\Phi(\a)\)
\end{align}
The polynomials $s_a$ constitute canonical basis with respect to
this pairing:
$$s_a\circ s _b=\text{sgn}(a)\delta _{a, -b}$$

For the polynomials of $e^{\a}$ one also introduces the pairing:
\begin{align}
P\circ Q=\int\limits _{-\infty}^{\infty} d\a
\frac{P(e^{\a})Q(-e^{\a})-P(-e^{\a})Q(e^{\a})}
{\prod (e^{2\a}+e^{2\b _j})} e^{2\a}
\end{align}
It is not difficult to give explicit formulae for canonical basis $S_a$
($|j|=1,\cdots , n-1$) satisfying
$$S_a\circ S_b=\text{sgn}(a)\delta _{a,-b}$$
but we shall not need them. Notice that the structure of the
pairing implies that $S_{-k}$ and $S_k$ should be taken as respectively
odd and even polynomials. They contain quasi-constants (symmetric
functions of $e^{\b _j}$ as coefficients). We can take
$$S_{-k}= e^{(2k-1)\a},\quad k=1,\cdots ,n-1$$
as half-basis.

The main property of deformed hyper-elliptic integrals is deformed
Riemann bilinear relation:
\begin{align}
&\sum\limits _{k=1}^{n-1}
\left( \langle S_k\ |\ s_a\rangle \langle S_{-k}\ |\ s_b\rangle-
\langle S_k\ |\ s_b\rangle \langle S_{-k}\ |\ s_a\rangle\right)=
\text{sgn}(a)\delta _{a, -b}\non\\
&\sum\limits _{k=1}^{n-1}\left( \langle S_a\ |
\ s_k\rangle \langle S_b\ |\ s_{-k}\rangle-
\langle S_b\ |\ s_k\rangle \langle S_a\ |\ s_{-k}\rangle\right)=\text{sgn}(a)
\delta _{a, -b}\non
\end{align}
This relation introduces into the game the symplectic
group $Sp(2n-2)$.
To finish this section let us write two more formulae following from
(\ref{gener}). First,
\begin{align}
&c(\a _1,\a _2)\equiv\sum\limits _{k=1}^{n-1}\(s_k(\a _1)s_{-k}(\a _2)-
s_k(\a _2)s_{-k}(\a _1)\)=\non\\
&=\frac {\prod\limits _{j=1}^{2n}
(\a _1 -\b _j+\frac {\pi i}2)}{\a _1-\a _2+\pi i}-
\frac {\prod\limits _{j=1}^{2n}
(\a _1-\b _j-\frac {\pi i}2)}{\a _1-\a _2-\pi i}-
\frac {\prod\limits _{j=1}^{2n}
(\a _2 -\b _j+\frac {\pi i}2)}{\a _2-\a _1+\pi i}+
\frac {\prod\limits _{j=1}^{2n}
(\a _2-\b _j-\frac {\pi i}2)}{\a _2-\a _1-\pi i}\non
\end{align}
Second, it is obvious from (\ref{gener}) that for any partition there is
a symmetric matrix $c _{kl}$ depending polynomially on rapidities such that
\begin{align}
&A_{k}(\a|\b _{i_1},\cdots ,\b_{i_n}|\b _{j_1},\cdots ,\b_{j_n})=\non\\
&=s_k(\a )+\sum\limits _{l=1}^{n-1}c _{kl}
(\b _{i_1},\cdots ,\b_{i_n}|\b _{j_1},\cdots ,\b_{j_n})s_{-l}(\a)\label{As}
\end{align}
which means that for any partition
the polynomials
$A_k(\a|\b _{i_1},\cdots ,\b_{i_n}|\b _{j_1},\cdots ,\b_{j_n})$, $s_{-k}(\a )$
constitute a canonical basis with respect to the above pairing.
$$A_k\circ A_l=s_{-k}\circ s_{-l}=0,\quad A_k\circ s_{-l}=\delta _{k,l}$$
\section{Level -4 from level 0.}

It is obvious from equations (\ref{symm}, \ref{Rie}) and
(\ref{symm0}, \ref{Rie0}) that for
any pair of solutions the scalar product
$$f(\b _1,\cdots , \b _{2n})g(\b _1, \cdots ,\b _{2n})$$
is a quasi-constant. So, if we manage to have a complete set of
solutions on level 0 the level -4 solutions are obtained by inverting the
square matrix. This is the main idea of our construction. Let us count
the solutions on level 0.

Consider the space of skew-symmetric polynomials of variables
$\a _1, \cdots ,\a _{n-1}$ with the basis
$\text{det}\| s_{a_p}(\a _q)\|_{p,q}$.
The group $Sp(2n-2)$ acts in this space as on the space of skew-symmetric
tensors.
In this space we define a subspace $H_{\text{irrep}}$ of maximal
irreducible representation of $Sp(2n-2)$ which is the orbit
of this group obtained by action on the polynomial
$\text{det}\| s_{p}(\a _q)\|_{p,q}$
This is the fundamental representation of maximal dimension:
$$ d_{\text{irrep}}=\binom{2n-2}{n-1}-\binom{2n-2}{n-3}$$
The formula (\ref{As}) implies that for any partition the polynomial
$$\text{det}(A_{k}(\a _l|\b _{i_1},\cdots ,\b_{i_n}|
\b _{j_1},\cdots ,\b_{j_n}))|_
{1\le k,l\le n-1}\in H_{\text{irrep}}$$
belongs to the representation of $Sp(2n-2)$ in skew-symmetric
tensors of degree $n-1$.

Due to deformed Riemann bilinear relation among the solutions
counted by $\{k_1,\cdots ,k_{n-1}\}$ only $d_{\text{irrep}}$ are linearly
independent over the ring of quasi-constants. To obtain them we take
the polynomial
$$\text{det}(S_{-k}(e^{\a _l}))$$
as basic one and obtain the rest as orbit under the action of Borel
subgroup, i.e. by
the matrices
$$
\begin{pmatrix}I &Z\\ 0 &I\end{pmatrix}\in Sp(2n-2)
$$
where the matrix $Z$ is symmetric. Thus we obtain $d_{\text{irrep}}$ linearly
independent solutions.

On the other hand the covectors $f(\b _1,\cdots ,\b _{2n})$ belong to
singlet subspace of $(\mathbb{C}^2)^{\otimes 2n}$. The dimension of
this subspace equals:
$$d_{\text{sing}}=\binom{2n}{n}-\binom{2n}{n-1}$$
The marvelous identity \cite{count}
$$d_{\text{irrep}}=d_{\text{sing}}$$
shows that we have exactly the same number of solutions as the dimension
of space. So, different linear independent solutions can be combined into
the square matrix
$\mathcal{F}(\b_1,\cdots ,\b _{2n})$. Now we can find the solutions on level -4
solving the equation:
$$ \mathcal{F}(\b_1,\cdots ,\b _{2n})\mathcal{G}(\b_1,\cdots ,\b _{2n})=I$$
So, our goal is to find an efficient way for inverting
the matrix $\mathcal{F}$.

Notice that $ \mathcal{F}(\b_1,\cdots ,\b _{2n})$  naturally splits into
the product:
$$\mathcal{P}(\b_1,\cdots ,\b _{2n})\mathcal{H}(\b_1,\cdots ,\b _{2n})$$
where the multipliers $\mathcal{P}$ and $\mathcal{H}$ carrying
respectively transcendental and rational dependence on $\b _j$ are
defined as follows.
\begin{align}
\mathcal{P}(\b_1,\cdots ,\b _{2n})=P_{\text{irrep}}
\widetilde{\mathcal{P}}(\b_1,\cdots ,\b _{2n})P_{\text{irrep}}\non
\end{align}
where $P_{\text{irrep}}$ is the projector on the irreducible representation
of $Sp(2n-2)$ discussed above and the matrix $\widetilde{\mathcal{P}}$
acts in the $(n-1)$-th skew-symmetric power of $\mathbb{C}^{2n-2}$,
its matrix elements are given by
$$ \text{det}(\langle S_{a_k}|s_{b_l}\rangle)|_{1\le k,l \le n-1}$$
The rational in $\b _j$ matrix $\mathcal{H}(\b_1,\cdots ,\b _{2n})$ acts
from the singlet subspace of $(\mathbb{C}^2)^{\otimes 2n}$
into the space of maximal irreducible representation
of $Sp(2n-2) $ in $\wedge ^{n-1}\mathbb{C}^{2n-2}$.
In the space $(\mathbb{C}^2)^{\otimes 2n}$
we take the basis $w_{\epsilon _1,\cdots,\epsilon _{2n}}(\b _1,\cdots\b_{2n})$
in such a way that the components of covectors
in this space are counted by partitions
$\b _{i_1},\cdots ,\b_{i_n}|\b _{j_1},\cdots ,\b_{j_n}$. Different vectors from
$\wedge ^{n-1}\mathbb{C}^{2n-2}$ are counted by
$-(n-1)\le a_1<a_2<\cdots < a_{n-1}\le (n-1)$. In this
basis the matrix elements of $\mathcal{H}(\b_1,\cdots ,\b _{2n})$
are
$$\frac 1 {\prod (\b _{i_p}-\b _{j_q})}
\text{det }(\widetilde{c}_{k,a_l}
(\b _{i_1},\cdots ,\b_{i_n}|\b _{j_1},\cdots ,\b_{j_n}))
|_{1\le k,l\le n-1}$$
where
$$\widetilde{c}=\begin{pmatrix} I &c\\ 0&I\end{pmatrix}$$
and ${c}_{i,j}$ is defined in
(\ref{As}).

Now we want to invert these matrices. Due to deformed Riemann
bilinear relation inverting of the transcendental part is trivial:
\begin{align}
\mathcal{P}(\b_1,\cdots ,\b _{2n})^{-1}=P_{\text{irrep}}
\widetilde{\mathcal{P}}^{\dag}(\b_1,\cdots ,\b _{2n})P_{\text{irrep}}\non
\end{align}
where the matrix elements of $\widetilde{\mathcal{P}}^{\dag}$ are given by
$$ \text{det}(\langle S_{b_k}^{\dag}|s_{a_l}\rangle)|_{1\le k,l \le n-1}$$
with
$$ S_{b}^{\dag} = \text{sgn}(b)S_{-b}$$
So, surprisingly enough the main difficulty happens to be in inverting
of the rational matrix. In this section we give one approach to
the problem which proves that the inverse matrix possesses
nice properties.

First, we have to take care of the basis $w$. What we actually need is
a construction of dual basis. This construction can be found in
\cite{book} , we do not give it explicitly here, the main properties
of the dual basis $w^{\dag}$ are:
\begin{align}
&\overline R(\b _{i+1}-\b _i)
w^{\dag}(\b _1, \cdots ,\b _i,\b _{i+1},\cdots ,\b _{2n})
_{\epsilon _1, \cdots ,\epsilon _i,\epsilon _{i+1},\cdots ,\epsilon_{2n}}
=\non\\&=
w^{\dag}(\b _1, \cdots ,\b _{i+1},\b _i,\cdots ,\b _{2n})
_{\epsilon _1, \cdots ,\epsilon _{i+1},\epsilon _i,\cdots ,\epsilon_{2n}}
,\non\\
&w(\b _1, \cdots ,\b _{2n})_
{\epsilon _1, \cdots ,
\epsilon_{2n}}w^{\dag}(\b _1, \cdots ,\b _{2n})_
{\epsilon _1', \cdots ,\epsilon_{2n}'}=
\prod \delta _{\epsilon _i,\epsilon _i'}
\non
\end{align}
Consider the operator $\mathcal{H}^*(\b _1,\cdots ,\b _{2n})$
which coincide with Hermitian conjugation of
$\mathcal{H}(\b _1,\cdots ,\b _{2n})$ for real $\b _j$ and then
is continued analytically. For the matrix elements of this
operator in usual basis for $\wedge ^{n-1}\mathbb{C}^{2n-2}$
and the basis $w^{\dag}$ in
$(\mathbb{C}^2)^{\otimes 2n}$ one finds \cite{book}:
$$\frac 1 {\prod (\b _{i_p}-\b _{j_q}+i\pi)}
\text{det }
(\widetilde{c}_{k,a_l}(\b _{i_1},\cdots ,\b_{i_n}|\b _{j_1},\cdots ,\b_{j_n}))
|_{1\le k,l\le n-1}$$
Let us write the identity:
$$\mathcal{H}^{-1}(\b _1,\cdots ,\b _{2n})=
\mathcal{H}^{*}(\b _1,\cdots ,\b _{2n})
\(\mathcal{H}(\b _1,\cdots ,\b _{2n})
\mathcal{H}^{*}(\b _1,\cdots ,\b _{2n})\)^{-1}
$$
The operator $\mathcal{H}\mathcal{H}^{*}$ is nicer
then $\mathcal{H}$ itself because it acts from the space
of irreducible representation of $Sp(2n-2)$ to itself.
Its matrix elements are:
\begin{align}
&\sum\limits
_{\{1,\cdots,2n\}=\{i_1,\cdots ,i_n\}\cup \{j_{1},\cdots ,j_{n}\}}
\frac 1 {\prod (\b _{i_p}-\b _{j_q}+i\pi)(\b _{i_p}-\b _{j_q})}\non\\
&\times
\text{det }(\widetilde{c}_{k,a_l}
(\b _{i_1},\cdots ,\b_{i_n}|\b _{j_1},\cdots ,\b_{j_n}))
\text{det }(\widetilde{c}_{k,b_l}
(\b _{i_1},\cdots ,\b_{i_n}|\b _{j_1},\cdots ,\b_{j_n}))\non
\end{align}
Unfortunately we could not find a way of efficiently inverting
this operator, but we were able to calculate its determinant:
\begin{align}
\text{det}(\mathcal{H}&(\b _1,\cdots ,\b _{2n})
\mathcal{H}^{*}(\b _1,\cdots ,\b _{2n}))=\non\\&=Const
\(\prod\limits _{i,j}(\b _i-\b _j-\pi i)\)^{-\(\binom{2n-4}{n-2}-
\binom{2n-4}{n-4}\)}\label{det}
\end{align}
The proof of this formula is based on two facts. First, one can easily
calculate the degree of the determinant as function of $\b $'s.
Second, the polynomial $u$ satisfies the following recurrence relation
\cite{book} :
\begin{align}
&u(\a _1,\cdots ,\a _{n-1}|\b_1,\cdots ,\b _{n-1},\b -\frac{\pi i}2
|\b _{n+1},\cdots ,\b _{2n-1},\b +\frac{\pi i}2)=\non\\&=
\prod_{j=1}^{n-1}(\a _j-\b )
\sum\limits _{j=1}^{n-1}(-1)^j
\(\prod\limits _k
(\a_j-\b_k +\frac {\pi i}2)-\prod\limits _k (\a_j-\b_k -\frac {\pi i}2)\)
\non\\&\times
u(\a _1,\cdots ,\widehat{\a _j},\cdots ,\a _{n-1}|\b_1,\cdots ,\b _{n-1}
|\b _{n+1},\cdots ,\b _{2n-1})
&
\label{reu}
\end{align}
Using this relations one can calculate the rank of residue of
$\mathcal{H}\mathcal{H}^*$ at the point $\b _i=\b _j +\pi i$.

Putting all this information together we arrive at the
following
\newline
{\bf Theorem.} {\it The solutions to qKZ equations on level -4
counted by $\{k_1,\cdots ,k_{n-1}\}$, with $|k_j|\le n-1$, $\forall j$
can be written in the following form:}
\begin{align}
&g^{\{k_1,\cdots ,k_{n-1}\}}(\b _1,\cdots, \b _{2n})=
\prod\limits _{i<j}\frac 1 {\zeta (\b _i-\b _j)}
\int\limits _{-\infty}^{\infty} d\a _1\cdots \int\limits _{-\infty}^{\infty}
d\a _{n-1}\prod\limits _{i,j}\varphi (\a _i -\b _j)\non\\
&\times
\text{det}|e^{k_i\a _j}|_{1\le i,j\le n-1}\
\ \widetilde{h}(\a _1,\cdots \a _{n-1}|\b _1,\cdots ,\b _{2n})
\label{theor}
\end{align}
{\it where $\widetilde{h}(\a _1,\cdots \a _{n-1}|\b _1,\cdots ,\b _{2n})$
is a polynomial of all its argument, skew-symmetric with
respect to $\a _1,\cdots \a _{n-1}$}
\newline
{\it Proof. }  The only point which remains to be proved is that
$\widetilde{h}$ is indeed a polynomial because {\it a priori} we
can be sure only that it is a rational function. The structure
of $h$ and the formula for determinant (\ref{det}) imply that
there are no other possible singularities than
poles at $\b _i=\b _j+\pi i$. By recurrence relation  following
from (\ref{reu}) the residue of $h$ at $\b _i=\b _j+\pi i$ is defined
by the same function $h$ for $n\to n-1$. So, the rank of the
residue is defined by the dimension of singlet subspace
of $\mathbb{C}^{2n-2}$:
$$\binom{2n-2}{n-1}-\binom{2n-2}{n-2}$$
which is the same as the exponent in (\ref{det}).
Now it is clear that in inverse
matrix the pole is canceled by zero coming from the determinant.
\hfill{\bf QED}

\vskip 0.3cm
As in level 0 case there is a linear dependence
between the solutions which is removed by Riemann bilinear
relation.

The problem of direct inverting the matrix $\mathcal{H}$ seems
to be too complicated. So, we need alternative ways to define
the polynomials $\widetilde{h}$. First, let us reformulate the
original definition. We can present $\widetilde{h}$ in the
following form:
\begin{align}
&\widetilde{h}(\a _1, \cdots ,\a _{n-1}|\b _1,\cdots ,\b _{2n})
=\non\\&=
\sum\limits
_{\{1,\cdots,2n\}=\{i_1,\cdots ,i_n\}\cup \{j_{1},\cdots ,j_{n}\}}
v(\a _1,\cdots \a _{n-1}|
\b _{i_1},\cdots ,\b _{i_n}|\b _{j_1},\cdots ,\b _{j_n})
\non\\&\times\prod\limits _{p,q=1}^n \frac {\b _{i_p}-\b _{j_q}+\pi i}
{\b _{i_p}-\b _{j_q}}
\ w^{\dag}_{\epsilon _1,\cdots,\epsilon _{2n}}(\b _1,\cdots\b_{2n})\non
\end{align}
Then the function $v$ is subject to two requirements. The first follows
from the fact that $\widetilde{h}$ must belong to singlet subspace.
Using the transformation of the basis $w^{\dag}$ under the action of $su(2)$
described in \cite{book} one finds the equations:
\begin{align}
&\sum\limits _{p=1}^{n+1}v(\a _1,\cdots ,\a _{n-1}|
\b _{i_1},\cdots ,\b _{i_{n-1}},\b _{j_p}|\b_{j_1},
\cdots ,\widehat {\b  _{j_p}},\cdots ,\b _{j_{n+1}})
\non\\&\times \prod\limits
_{q\ne p}\frac {\b _{j_p}-\b _{j_q} -\pi i}{\b _{j_p}-\b _{j_q}}=
0\label{singlet}
\end{align}
The second equation is equivalent to the fact that $\widetilde{h}$
is obtained by inverting the matrix $\mathcal{H}$:
\begin{align}
&\sum\limits
_{\{1,\cdots,2n\}=\{i_1,\cdots ,i_n\}\cup \{j_{1},\cdots ,j_{n}\}}
v(\a _1,\cdots ,\a _{n-1}|
\b _{j_1},\cdots ,\b _{j_n}|\b _{i_1},\cdots ,\b _{i_n})
\non\\
&
\times u(\a _1',\cdots ,\a _{n-1}'|
\b _{i_1},\cdots ,\b _{i_n}|
\b _{j_1},\cdots ,\b _{j_n})\prod
\limits _{p,q=1}^n\frac 1 {\b _{i_p}-\b _{j_q}}=\non\\&=
c(\a _1,\cdots ,\a _{n-1}|\a _1',\cdots ,\a _{n-1}')\label{intersect}
\end{align}
where $c(\a _1,\cdots \a _{n-1}|\a _1',\cdots \a _{n-1}')$
is the "intersection form". Essential part of this "intersection form" is
$\text{det}\left|c(\a _i,\a '_j)\right|$, but some
additional terms should be added in order that
$c(\a _1,\cdots \a _{n-1}|\a _1',\cdots \a _{n-1}')$
belongs to $H_{\text{irrep}}$
with respect to both sets $\a _1,\cdots \a _{n-1}$ and
$\a _1',\cdots \a _{n-1}'$.
Introduce Grassmann variables $\xi _j $, $\eta _j$ ($j=1,\cdots , n-1$):
$$\xi _i\xi _j=-\xi _j\xi _j,\quad \xi _i\eta _j=-\eta _j\xi _j,\quad
\eta _i\eta _j=-\eta _j\eta _i$$
Let
\begin{align}
&C=\sum\limits _{i,j} c(\a _i,\a _j')\xi _i\eta _j,\non\\
&S=\sum\limits _{i<j} c(\a _i,\a _j)\xi _i\xi _j,\non\\
&S'=\sum\limits _{i<j} c(\a _i',\a _j')\eta _i\eta _j,\non
\end{align}
Then
\begin{align}
&c(\a _1,\cdots \a _{n-1}|\a _1',\cdots \a _{n-1}')\xi _1\cdots \xi _{n-1}
\eta _1\cdots \eta _{n-1}=
\sum\limits _{k=0}^{\[\frac {n-1} 2\]} c _k(SS')^kC^{n-1-2k}\non
\end{align}
where the coefficients $c_k$ are
$$c_0=1,\quad c_k=\frac{(n-1)!}{k!(k+1)!(n-2k-1)!}$$
If we consider
$v(\a _1,\cdots \a _{n-1}|
\b _{i_1},\cdots ,\b _{i_n}|\b _{j_1},\cdots ,\b _{j_n})$
for different partitions as $ \binom{2n}{n} $
independent unknowns then we have sufficient number of linear equations:
$ \binom{2n}{n}-\binom{2n}{n-1}$ from (\ref{intersect}) and $\binom{2n}{n-1}$
from (\ref{singlet}).

Our main goal is to describe efficiently
the polynomials $\widetilde{h}$.
The way to approach this problem will be discussed in another paper.
To finish the present paper we
would like to give an intuitive idea about reasons behind
the possibility of rewriting original Jimbo-Miwa formula
in the form without denominators.

\section{Cohomological meaning of new formula.}

Consider the "classical" limit:
$$\b _j =\frac 1 {\hbar}x _j, \quad \a =\frac 1 {\hbar}z\qquad \hbar\to 0$$
In this limit
\begin{align}
&\langle P\ |\ p\rangle=
\int\limits _{-\infty}^{\infty}
\prod\limits _j\varphi (\a -\b _j)\ P(e^{\a })\ p(\a )
\ e^{-(n-1)\a }d\a 
\to \int\limits _{\gamma}\frac {p(z)} wdz\non
\end{align}
where the hyper-elliptic surface $X$ is defined by
$$w^2=\prod (z-x_j),$$
There are two points ($\infty ^{\pm}$) on the curve lying
above the point $z=\infty$.
The genus equals $n-1$.
The contour $\gamma$ is defined by $P$. In particular,
$$S_{-k}\leftrightarrow b_k, \quad S_{k}\leftrightarrow a_k$$
The polynomials
$$\widetilde{s}_a=\lim _{\hbar\to 0} s_a$$
describe canonical basis of differentials.
Namely for
$$\sigma _a=\frac {\widetilde{s}_a (z)} {w}dz$$
one has
$$\sigma _a\circ \sigma _b =\sum\limits _{\infty ^{\pm}}
\text{res}\(\sigma _a d^{-1}(\sigma _b)\)=\text{sgn}(a)\delta _{a, -b}$$
The differentials $\sigma _k$ are of first kind, $\sigma _{-k}$ -
of second, there is also the third kind differential
$$\sigma _0=\frac {z^{n-1}}w dz$$
Consider the Jacobi variety of $X$:
$$J=\mathbb{C}^{2n-2}/(\mathbb{Z}^{n-1}+B\mathbb{Z}^{n-1})$$
where $B$ is the matrix of $B$-periods of normalized holomorphic
differentials:
$$B_{ij}=\int\limits _{b_i}\omega _j$$
We define Riemann
theta-function
$$\theta (\zeta), \qquad \text{for}\quad
\zeta \in \mathbb{C}^{2n-2}$$

Consider the divisor $\{P_1,\cdots ,P_g\}$ where $P_j$ are points on the Riemann
surface: $P_j=\{z_j,w_j\}$.
Abel transformation is defined as follows
$$\{P_1,\cdots , P_{n-1}\}\to \zeta =
\sum\int\limits ^{P_j}\omega$$
defines map
$$\text{Symm}(X^{n-1})\to J$$
which is not one to one. However, if we consider
non-compact varieties
$$J-(\Theta _-\cup\Theta _+)$$
where
$$\Theta _{\pm}=\{\zeta|\theta (\zeta
+\rho_{\pm})=0\},\ \rho _{\pm}=\int\limits ^{\infty ^{\pm}}\omega$$
and
$$\text{Symm}(X^{n-1})-D$$
where
$$D=\{\{P_1,\cdots , P_{n-1}\}|P_j=\infty ^{\pm}, P_i=\sigma(P_j)\}$$
they are isomorphic. The integrand of the Jimbo-Miwa
formula gives in the classical limit a differential form on
$$\text{Symm}(X^{n-1})-D$$
which is isomorphic to
$$f(\zeta)d\zeta _1\cdots d\zeta _{n-1}$$
with $f(\zeta)$ meromorphic on $J$ with poles on $\Theta _{\pm}$.
The question arises concerning cohomologies. They are described
by the following theorem conjectured in \cite{ns} and proved
in \cite{n}.
\newline
{\bf Theorem} (A. Nakayashiki) {\it The dimension of
cohomologies space is }
$$\binom{2n-1}{n-1}-\binom{2n-1}{n-3}$$
{\it in terms of $\text{Symm}(X^{n-1})-D$ it is realized as follows.
$2n-1$ differential of 1-st, 2-ond, 3-d kind $\sigma _a$. Let}
$$\widetilde{\sigma}_a=\sum\limits _{k=1}^{n-1}{\sigma}_a(P _k)$$
{The cohomologies are realized as follows:}
$$\{\widetilde{\sigma}_{a_1}\wedge\cdots \wedge
\widetilde{\sigma }_{a_{n-1}}\}/
\{\omega\wedge\widetilde{\sigma}_{a_1}\wedge\cdots
\wedge \widetilde{\sigma}_{a_{n-3}}\}$$
{\it where}
$$\omega =\sum\limits _{k=1}^{n-1}
\widetilde{\sigma}_{k}\wedge\widetilde{\sigma}_{-k}$$

This is the reason why classical limit of Jimbo-Miwa formula
can be reduced to one-fold integrals. We suppose that similar
interpretation is possible in deformed case.

\section{Conclusion}

In this communication
the problem to reduce the Jimbo-Miwa solution of the qKZ
on level -4 to one-fold integrals is solved only partially.
As we have shown it is related to the cohomological
origin of our formulae.
In our next paper we shall give an explicit form of the
polynomials $\widetilde h$ from the formula (\ref{theor}).
We shall also discuss the relation of the above
solution to the correlation functions of the XXX model.
In order to treat this problem properly we need to carry out
an accurate analysis
of singularities which appear in intermediate stage. The main task is
to prove that final result for the correlation functions is
really finite. The explicit form of this result is in agreement with
the ansatz from the paper \cite{bks}. The conjecture \cite{bk1,bk2}
about the structure of the correlation functions in the homogeneous limit
through the Riemann $\zeta$-function at odd argument follows from
the form of the correlation functions in the inhomogeneous case
which in it's turn follows from the solution to the qKZ on level -4.

\section{Acknowledgements}

The authors would like to thank R. Flume,
M. Jimbo and T. Miwa for useful discussions.
This research  has been supported by the following grants:
NSF grant PHY-9988566, the Russian Foundation of Basic Research
under grant \# 01--01--00201, by INTAS under grants \#00-00055 and \# 00-00561.
HEB would like to thank the administration of
the Max-Planck Institute for Mathematics
for hospitality and perfect conditions for the work.

\end{document}